\documentclass[10pt,a4paper,twoside]{article}
\usepackage{epsfig}
\usepackage{baltlat6}
\usepackage{array}
\usepackage{here}
\begin{document}
 \ \ \vspace{0.5mm} \setcounter{page}{0}

 \titlehead{Baltic Astronomy, vol.\,99, 999--999, 2016}
 \titleb
 {DETERMINING THE GALACTIC BAR PARAMETERS
 BASED ON THE HERCULES AND WOLF\,630 STELLAR STREAMS}

\begin{authorl}
\authorb{A.T. Bajkova}{} and
\authorb{V.V. Bobylev}{}
\end{authorl}

\begin{addressl}
 \addressb{}{Central (Pulkovo) Astronomical Observatory of RAS, 65/1 Pulkovskoye Ch.,
 St. Petersburg, Russia; anisabajkova@rambler.ru}

\end{addressl}

\submitb{Received: 2016 June 99; accepted: 2016 June 99}

\begin{summary}
We have identified the four most significant features in the $UV$
velocity distribution of solar neighborhood stars: H1,H2 and W1,W2
in the Hercules and Wolf\,630 streams respectively. We have put
the task of determining number of characteristics of the central
Galactic bar independently from each of the identified features by
assuming that the Hercules and Wolf\,630 streams have a
bar-induced dynamical nature. The problem has been solved by
constructing $2:1$ resonant orbits in the rotating bar frame for
each star in these streams. Analysis of the resonant orbits found
has shown that the bar pattern speed $\Omega_{b}$ lies within the
range 45--52 km s$^{-1}$ kpc$^{-1}$ with a mean of $48.1\pm1.0$ km
s$^{-1}$ kpc$^{-1}$, while the bar angle $\theta_{b}$ is within
the range $35^\circ-65^\circ$ with a mean of $50\pm4^\circ.$ The
results obtained are consistent with the view that the Hercules
and Wolf\,630 streams could be formed by a single mechanism
associated with the fact that a long-term influence of the
Galactic bar led to a characteristic bimodal splitting of the $UV$
velocity plane.
 \end{summary}

 \begin{keywords}
 Kinematics -- Stellar Streams -- Hercules Stream -- Wolf\,630 Stream -- Galactic Bar -- Galaxy (Milky Way)
 \end{keywords}

 \resthead
 {Determining the Galactic bar parameters}
 {A.T. Bajkova, V.V. Bobylev}

\sectionb{1}{INTRODUCTION}
Analysis of the velocity field of solar-neighborhood stars using
Hipparcos (1997) data (Chereul et al. 1998; Dehnen 1998; Asiain et
al. 1999; Skuljan et al. 1999) and based on the most recent data
(Famaey et al. 2005; Bobylev \& Bajkova 2007; Antoja et al. 2008;
Bobylev et al. 2010) has revealed a well-developed fine structure.
Various nonaxisymmetric Galactic potential models, in particular,
the spiral pattern and the Galactic bar, are considered to explain
several structures to which quite old stars belong.

As simulations showed, the existence of the Hercules stream can be
explained by the fact that its stars have resonant orbits induced
by the Galactic bar (Dehnen 1999, 2000; Fux 2001; Chakrabarty
2007; Gardner \& Flinn 2010). A detailed analysis of the kinematics of nearby F and G
dwarfs using high-resolution spectra (Bensby et al. 2007) showed
the stars in the stream to have a wide range of stellar ages,
metallicities, and elemental abundances, and this led to the
conclusion that the dynamical effect of the Galactic bar is the
most acceptable explanation for the existence of the Hercules
stream.

In addition to the Hercules stream, there are also other features
on the $UV$ velocity plane, in particular, the Wolf\,630 stream
(Bobylev et al. 2010), with some of the authors designating this
place as the $\alpha$~Ceti stream (Francis \& Anderson 2009). The
interest is that there are orbits both elongated along the bar
major axis and oriented perpendicularly to it in the bar reference
frame. At present, there is reason to believe that the Sun is near
the point of intersection of such orbits (Fig.~1 in Dehnen 2000).
In this case, the bimodality of the $UV$ velocity distribution can
be explained by the fact that there are representatives of these
two orbit families in the solar neighborhood. According to Dehnen
(2000), the detailed form of the velocity distribution depends
strongly on the position of the Outer Lindblad Resonance (OLR). In
particular, there will be no bimodality if the OLR is farther from
the Galactic center than the solar cycle ($R_{OLR}/R_0>1.05,$
Fig.~4 in Dehnen (2000)). According to the coordinates on the $UV$ velocity plane, the
Hercules stream is a representative of the family of orbits
oriented perpendicularly to the bar major axis, while the
representatives of the family of orbits elongated along the bar
major axis are located in the region on the $UV$ velocity plane
where the Wolf\,630 stream is observed.

The goal of this paper is to estimate such characteristics of the
bar as its pattern speed $\Omega_{b}$ and orientation $\theta_{b}$
under the assumption that the Hercules and Wolf\,630 streams could
be formed by a single mechanism associated with the influence of
the Galactic bar. We solve the problem by constructing $2:1$
resonant orbits in the rotating bar frame. The characteristics of
each of the streams are determined independently, i.e., we do not
assume them to be of common origin in advance. This work differs
from the one published recently (Bobylev \& Bajkova 2016), in
which for stellar orbit construction we used equations of motion
in a coordinate system centered on the Sun and rotating around
Galactic center with a constant angular velocity (Fern\'andez et
al. 2008, Asiain et al. 1999). In this paper we use more exact
equations of motion in Galactic rotating frame based on effective
potential $\Phi_{eff}(x,y,z)$ (Jung \& Zotos 2016). As a result we
have obtained new, slightly different, estimates for the Galactic
bar parameters.

%%%%%%%%%%%%%%%%%%%%%%%%%%%%%%%%%%%%%%%%%%%%%%%%%%%%%%%%%%%%%%%%% Fig 1
 \begin{figure} {\begin{center}
 \includegraphics[width=60mm]{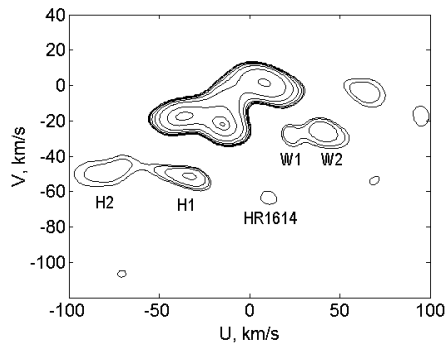}
 \caption{\small Wavelet map of $UV$ velocities constructed using $\sim$17000 single stars
 from Hipparcos (1997) with relative parallax errors of no more than 10\%.}
 \label{f1} \end{center} } \end{figure}
%%%%%%%%%%%%%%%%%%%%%%%%%%%%%%%%%%%%%%%%%%%%%%%%%%%%%%%%%%%%%%%%%

\sectionb{2}{THE UV VELOCITY DISTRIBUTION OF THE STREAMS}\label{UV}
The wavelet map of the $UV$ velocity distribution for single stars
with reliable distance estimates from Bobylev et al. (2010) is
presented in Fig.~1. The map was constructed using $\sim$17000
stars from the Hipparcos catalogue (1997). We took their proper
motions and parallaxes from a revised version of this catalogue
(van Leeuwen 2007) and their radial velocities from the OSACA
catalogue of radial velocities (Bobylev et al. 2006) and the
Pulkovo compilation of radial velocities (PCRV) (Gontcharov 2006).
For all these stars, the relative error in the parallax does not
exceed 10\%.

The contour lines in Fig. \ref{f1} are given on a logarithmic
scale: 1, 2, 4, 8, ..., 90\%, and 99\%. The velocities are given
relative to the Sun. This figure indicates the W1 and W2 features
for the Wolf\,630 stream and the H1 and H2 features for the
Hercules stream with the following coordinates of their centers:
W1 $(U,V)=(21,-26)$ km s$^{-1}$, W2 $(U,V)=(40,-24)$ km s$^{-1}$,
H1 $(U,V)=(-37,-50)$ km s$^{-1}$, H2 $(U,V)=(-78,-49)$ km
s$^{-1}$. According to this map, we selected the specific stars
belonging to these features based on the probabilistic method. The
number of probable candidates in each stream was the following:
250, 271, 401, and 218 stars in W1, W2, H1, and H2, respectively.

When constructing the Galactic orbits to determine the stellar
velocities relative to the local standard of rest (LSR), we use
the peculiar solar velocity components relative to the LSR with
their values from Sch\"onrich et al. (2010), $(U_\odot,
V_\odot,W_\odot)=(11.1, 12.2, 7.3)\pm(0.7, 0.5, 0.4)$ km s$^{-1}$.

\sectionb{3}{METHOD}\label{Method}
\subsectionb{3.1}{Orbit construction}
The Galactic bar follows a clockwise rotation around the z-axis at
a constant angular velocity $\Omega_b$. The potential in the
rotating frame of reference (known as the effective potential) is (Jung \& Zotos 2016)
\begin{equation}
\Phi_{eff}(x,y,z)=\Phi_t(x,y,z)-\frac{1}{2}\Omega_b^2(x^2+y^2),
\label{Eff}
\end{equation}
where $\Phi_t(x,y,z)$ is the total Galactic potential, the
coordinate system $(x,y,z)$ is centered on the Galactic center,
with the $x$ axis being directed to the Galactic center, the $y$
axis pointing in the direction of Galactic rotation, and the
$z$ axis being directed toward the north Galactic pole.

The Hamiltonian which governs the motion of a test particle with a
unit mass  in our rotating galaxy model is
\begin{equation}
H=\frac{1}{2}(p_x^2+p_y^2+p_z^2)+\Phi_t(x,y,z)-\Omega_b L_z=E,
 \label{Ham}
\end{equation}
where $p_x$, $p_y$ and $p_z$ are the canonical momenta per unit
mass, conjugate to $x$, $y$ and $z$ respectively, $E$ is the
numerical value of the Jacobi integral, which is conserved, while
$L_z = x p_y - y p_x$.

The corresponding equations of motion are
\begin{equation}
 \begin{array}{llllll}
 \dot{x}=p_x+\Omega_b y,\\
 \dot{y}=p_y-\Omega_b x,\\
 \dot{z}=p_z,\\
 \dot{p}_x=-\partial\Phi_t/\partial x+\Omega_b p_y,\\
 \dot{p}_y=-\partial\Phi_t/\partial y-\Omega_b p_x,\\
 \dot{p}_z=-\partial\Phi_t/\partial z,
 \label{motion}
 \end{array}
\end{equation}
where the dot indicates the derivative with respect to the time.
When the Galactic potential $\Phi_t$ is known, the system of
equations (\ref{motion})  can be solved numerically. We used a
fourth-order Runge–Kutta integrator.

The Galactic potential $\Phi_t(x,y,z)$, considered here, consists
of an axisymmetric component and a bar potential:
 \begin{equation}
 \Phi = \Phi_o +\Phi_b.
 \label{e0}
 \end{equation}

In turn, the axisymmetric component can be represented as the sum
of three components --- central (bulge), disk, and halo ones:
 \begin{equation}
 \Phi_o = \Phi_C + \Phi_D + \Phi_H.
 \label{e1}
 \end{equation}

We used the model of Allen \& Santill\'an (1991). The central
component of the Galactic potential is represented as
 \begin{equation}
 \Phi_C=-\frac{M_C}{(x^2+y^2+z^2+b^2_C)^{1/2}}.
 \label{2}
 \end{equation}
where $M_C$ is the mass, $b_C$ is the scale length, and
$r^2=x^2+y^2,$

The disk component is
 \begin{equation}
 \Phi_D=-\frac{M_D}{\{x^2+y^2+[a_D+(z^2+b_D^2)^{1/2}]^{2}\}^{1/2}},
 \label{3}
\end{equation}
where $M_D$ is the mass, $a_D$ and $b_D$ are the scale lengths.

The halo component is
 \begin{equation}
 \Phi_H=-\frac{M(R)}{R}-\int_R^{100}{{\frac{1}{R^{'}}}{\frac{dM(R^{'})}{dR^{'}}}}dR^{'},
 \label{4}
 \end{equation}
where
$$
M(R)=\frac{M_H(R/a_H)^{2.02}}{1+(R/a_H)^{1.02}}.
$$
Here, $M_H$ is the mass, $a_H$ is the scale length, and
$R^2=x^2+y^2+z^2.$

The triaxial ellipsoid model (Palou$\breve{s}$ et al. 1993) was
chosen as the potential due to the central bar:
\begin{equation}
  \Phi_b = \frac{M_b}{(q_b^2+x^2+[ya_b/b_b]^2+[za_b/c_b]^2)^{1/2}},
\label{bar}
\end{equation}
where $x=R\cos\vartheta, y=R\sin\vartheta$, $a_b, b_b, c_b$ are
the three bar semiaxes, $q_b$ is the bar length;
$\vartheta=\theta-\Omega_b t-\theta_b$, $tg(\theta)=y/x$;
$\Omega_{b}$ is the bar pattern speed, $t$ is the integration
time, $\theta_b$ is the bar angle relative to the Galactic $x$ and
$y$ axes.

The bar is introduced slowly. The total mass in the bar $M_b$
grows with the time
$$
M_b(t)=e^{-t_0/t}M_{b,final},
$$
where $t$ is the time, $t_0=0.5$ Gyr.

If $R$ is measured in kpc, $M_C,M_D,M_H, M_b$ are in units of the
Galactic mass $(M_G)$ equal to $2.325\times10^7M_\odot,$ then the
gravitational constant $G=1$ and 100 km$^2$ s$^{-2}$ is the unit
of measurement of the potential $\Phi_t$ and its individual
components (\ref{2})--(\ref{bar}). The parameters of the model
potential from Allen and Santill\'an (1991) and the bar potential
are:  $M_C=606 M_G, M_D=3690 M_G, M_H=4615 M_G, M_b=43.1 M_G$,
$b_C=0.3873$ kpc, $a_D=5.3178$ kpc, $b_D=0.25$ kpc, $a_H=12$ kpc,
$q_b=5$ kpc, $a_b/b_b=1/0.42$, $a_b/c_b=1/0.33$.

In the case of integration by the Runge--Kutta method, it should be
kept in mind that $t=0.001/1.023$ corresponds to one million years
at the distance and mass measurements adopted above. The orbit
integration time was chosen to be 2.2~Gyr, given that the
characteristic lifetime of the bar (from the beginning of its
formation to its destruction) is typically 2--4 Gyr.

\subsectionb{3.2}{Searching for Periodic Orbits}
This section is devoted to searching for resonant periodic orbits
in an axisymmetric potential in a rotating frame lying in the
Galactic plane centered at the Galactic center.

In this paper, as the rotating frame we consider the bar whose
pattern speed and orientation we attempt to estimate by
constructing the resonant periodic orbits of stars from the
Hercules and Wolf\,630 streams in the bar frame. This requires
finding the periodic orbits that satisfy the $(2:1)$ resonance
(Fux 2001). The condition for a resonance in the stellar disk
plane is the commensurability of two frequencies: the angular
velocity $\Omega$ and the epicyclic frequency $\kappa$:
 \begin{equation}
 l\kappa=m(\Omega-\Omega_p),
  \label{9}
 \end{equation}
with integers $m\ge 0$ and $l$. In the rotating frame, the $m/l$
resonant orbit is closed after $m$ radial oscillations and $|l|$
orbital periods. Note that condition (10) coincides with the
condition for Lindblad resonances only at $l=\pm1.$ The outside
corotation, $l<0,$ is negative, and we can talk about the outer
$m/|l|$ resonances. In this paper, we discuss only the outer
resonances. Fig.~4 from Fux (2001) presents such orbits at
$|l|=1.$ If $m=2$ and $l=-1,$ corresponding to $(2:1)$ resonant
orbits, then it follows from Eq.~(10) that
 \begin{equation}
 \Omega_p=\Omega+\kappa/2.
 \label{99}
 \end{equation}
The ideology of the proposed method is that we seek for the bar
pattern speed $\Omega_b$ (it is different for different stars) at
which the orbit becomes resonant and, hence, periodic (in the case
of an axisymmetric potential) and quasi-periodic (in the case of a
nonaxisymmetric potential, for example, when the bar is included,
see below). We propose the following numerical algorithm
seeking for such a pattern speed $\Omega_b$ lying within the range
$[\Omega_1,\Omega_2]$ predetermined empirically:

(1) For a given frequency range, we specify a grid of discrete
frequencies $\Omega_1+m\Delta\Omega, m=0,...,M$, where
$\Delta\Omega=(\Omega_2-\Omega_1)/M$, $M$ is an integer large
enough to obtain the required dense grid of frequencies.
Experience shows that an acceptable accuracy of the algorithm is
achieved at $\Delta\Omega=0.01$ km s$^{-1}$ kpc$^{-1}$.

(2) We integrate the stellar orbit equations (\ref{motion}) for a
sufficiently long time (several billion years or tens of
revolutions around the Galactic center) for each value of
$\Omega_b=\Omega_1+m\Delta\Omega, m=0,...,M$ specified above.

(3) We superimpose a discrete grid on the $(xy)$ plane with the
same step $\Delta x=\Delta y=\Delta$ in the $x$ and $y$ directions
that is chosen empirically to be small enough to provide the
required orbit reproduction accuracy. The product $\Delta \times
N$, where $N$ is the maximum number of cells in both $x$ and $y$
directions, gives the linear size of the $(xy)$ map.

(4) On the discrete $N\times N$ $(xy)$ plane, we assign ``1'' to
the coordinates of the cell through which the orbit passes and
``0'' to all the remaining discrete coordinates. If the orbit
crosses a given cell several times, then we anyway assign ``1'' to
the corresponding point on the plane only once.

(5) For all maps, we count the number of ``1''. The map with the
smallest number of ``1'' corresponds to the orbit with a resonance
frequency $\Omega_b=\Omega_1+i\Delta\Omega$, where $i$ is the map
number.

Note that this algorithm slightly differs from the one proposed by Bobylev \& Bajkova (2016)
because of another form of equations of motion.

Stable periodic orbits give a filling in the form of a trajectory
whose shape does not change with increasing number of revolutions
around the Galactic center. If several stable periodic orbits fall
within a given frequency range, then the algorithm chooses the
orbit with the smallest filling to which the orbit with the
smallest multiplicity of resonance frequencies corresponds. If the
orbit is not stable, then it progressively densely fills a certain
space on the $(xy)$ coordinate plane with increasing integration
time or number of revolutions around the Galactic center.

We also studied influence of the bar to the resonance orbits.
Naturally, introducing the bar leads in general to non-periodic (quasi-periodic)
orbits acquiring the property of stochasticity. But, it turned out that
when the bar was included, the resonance frequencies changed by a
negligibly small value, while the shape and orientation of orbits
in the bar frame practically did not change. Thus, we showed that
to solve our problem of determining the bar pattern speed and
orientation, it would suffice to study the periodic orbits
obtained in the axisymmetric potential. This is important, because
the bar potential is not known with certainty; there are a
multitude of its models in the literature. We only know that its
gravitational contribution to the total Galactic potential is
small.

The bar orientation in the Galactic $(xy)$ plane can be
determined from the orientation of resonant orbits in the bar
rotating frame. Indeed, while in the axisymmetric case the
orientation of the resonant orbits is arbitrary, the virtual
introduction of a bar will retain only those orbits which are
reflection symmetric with respect to at least one of the bar
principal axes (Fux 2001).

%%%%%%%%%%%%%%%%%%%%%%%%%%%%%%%%%%%%%%%%%%%%%%%%%%%%%%%%%%%%%%%%% Fig 2
 \begin{figure} {\begin{center}
 \includegraphics[width=70mm]{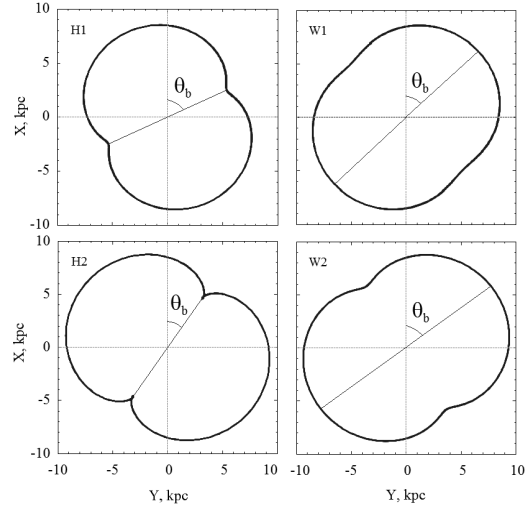}
 \caption{\small Resonant periodic orbits in the bar rotating frame constructed from averaged data on stars of H1, H2 and W1, W2 streams in axisymmetric Galactic potential. The Galactic center lies at
the coordinate origin; the Sun's coordinates are
$(x,y)=(8.5,0)$~kpc; $\theta_b$ is the bar angle counted from the direction to the Sun.}
 \label{f2} \end{center} } \end{figure}
%%%%%%%%%%%%%%%%%%%%%%%%%%%%%%%%%%%%%%%%%

%%%%%%%%%%%%%%%%%%%%%%%%%%%%%%%%%%%%%%%%%%%%%%%%%%%%%%%%%%%%%%%%% Fig 2
 \begin{figure} {\begin{center}
 \includegraphics[width=70mm]{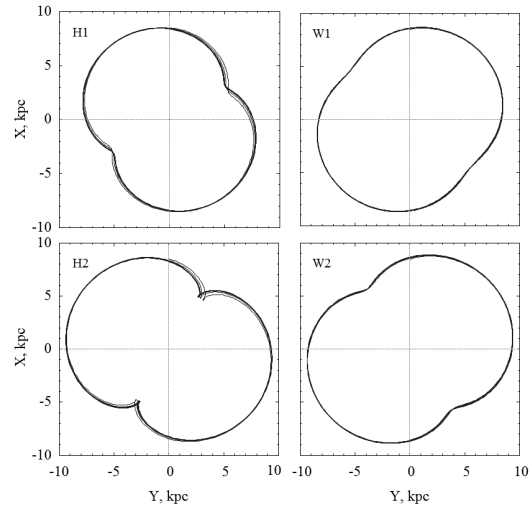}
 \caption{\small Resonant orbits in the bar rotating frame constructed from averaged data on stars of H1, H2 and W1, W2 streams in nonaxisymmetric Galactic potential (with the bar).}
 \label{f3} \end{center} } \end{figure}
%%%%%%%%%%%%%%%%%%%%%%%%%%%%%%%%%%%%%%%%%

%%%%%%%%%%%%%%%%%%%%%%%%%%%%%%%%%%%%%%%%%%%%%%%%
 {\begin{table}[t]                            %% T~2.
 \caption[]
 {\small\baselineskip=1.0ex
 Parameters found from the H1, H2 and W1, W2 features  }
 \label{t:2}
 \begin{center}\begin{tabular}{|c|c|c|}\hline
 Feature & $\Omega_{b},$ km s$^{-1}$ kpc$^{-1}$ & $\theta_{b},$ deg\\\hline
 $H_1$ & $51.4\pm0.8$ & $65\pm4$ \\
 $H_2$ & $49.1\pm1.4$ & $35\pm5~$ \\
 $W_1$ & $46.5\pm0.8$ & $45\pm3$ \\
 $W_2$ & $45.3\pm1.0$ & $55\pm4~$ \\\hline
 Mean & $48.1\pm1.0$ & $50\pm4~$ \\\hline
 \end{tabular}\end{center}\end{table}}
%%%%%%%%%%%%%%%%%%%%%%%%%%%%%%%%%%%%%%%%%%%%%%

\sectionb{4}{RESULTS AND DISCUSSION}
Using the above-described method of searching for periodic orbits
in an axisymmetric potential for each star of the identified H1,
H2, W1, and W2 features, we found the resonance frequencies
$\Omega_{2:1}$. The interval of
integration was 2.2~Gyr.

Fig. \ref{f2} presents the resonant periodic orbits in the bar rotating frame constructed from averaged data on coordinates and velocities of stars included in H1, H2 and W1, W2 streams using axisymmetric Galactic potential.
Fig. \ref{f3} shows the similar resonant orbits constructed using nonaxisymmetric Galactic potential which includes the potential of the bar. When the bar was included, the resonance frequencies changed by a very
small value (on average, 0.4\%) , while the shape and orientation of orbits
in the bar frame practically did not change. Therefore we determined parameters of the bar
from the periodic orbits (Fig.\ref{f2}).

As can be seen from Fig. \ref{f2}, the orbits corresponding to H1 and H2 streams are
elongated perpendicularly to the bar major axis, while the W1 and W2 orbits are
elongated along the bar, the bar angle $\theta_{b}$ lies within the range $35^\circ-65^\circ$.
In accordance with the central limit theorem, we assume that the deviations of the resonance
frequencies and the orbital inclinations of stars from the samples considered obey a
normal law. Both $\Omega_{b}$ and  $\theta_b$ for each sample were calculated as the
mean values of the individual parameters of stars from the samples; the
errors in $\Omega_{b}$ and  $\theta_b$ were calculated as the root--mean--square
deviation from the mean. The results obtained are presented in Table~1.

So, the resonant periodic orbits we found show that the bar pattern speed
$\Omega_{b}$ lies within the range 45--52 km s$^{-1}$
kpc$^{-1}$ with a mean of $48.1\pm1.0$ km s$^{-1}$ kpc$^{-1}$, while the bar
angle $\Omega_{b}$ is within the range $35^\circ-65^\circ$ with a
mean of $50\pm4^\circ.$ These parameters differ a little from the previously
obtained (compare with Table~2 from Bobylev \& Bajkova (2016)).

The results obtained are consistent with the view that the
Hercules and Wolf\,630 streams could be formed by a single
mechanism associated with the splitting of the $UV$ velocity plane
under the influence of the Galactic bar. Our results are in
agreement with the simulations of Dehnen (1999) for
$R_{OLR}\approx0.9R_0$, while in our case, as shown in Bobylev \&
Bajkova (2016), $R_{OLR}=0.93R_0$

\sectionb{5}{CONCLUSIONS}\label{conclusions}
We analyzed the four most significant features in the $UV$
velocity distribution of solar-neighborhood stars: H1, H2 from the
Hercules stream and W1, W2 from the Wolf\,630 stream with the
$(U,V)$ coordinates of their centers $(-37,-50), (-78,-49),
(21,-26),$ and $(40,-24)$ km s$^{-1}$, respectively. Based on the
assumption that the Hercules and Wolf\,630 streams were induced by
the central Galactic bar, we formulated the problem of determining
the bar characteristics independently from each of the identified
features.

To construct the Galactic orbits of the individual stars forming
these streams, we used the Galactic potential model of Allen \& Santill\'an (1991) and the equations of motion in Galactic rotating frame
based on effective potential $\Phi_{eff}(x,y,z)$ (Jung \& Zotos 2016).
From the set of constructed orbits, we
selected only the stable orbits in resonance with the bar.
Analysis of the resonant orbits found showed that the bar pattern
speed $\Omega_{b}$ lies within the range 45--52 km s$^{-1}$
kpc$^{-1}$ with a mean of $48.1\pm1.0$ km s$^{-1}$ kpc$^{-1}$,
while the bar angle $\theta_b$ is within the range $35^\circ-65^\circ$ with a
mean of $50\pm4^\circ.$ These estimates differ slightly from the previously
obtained (Bobylev \& Bajkova 2016).

The results obtained are consistent with the view that the Hercules
and Wolf\,630 streams could be formed by a single mechanism
associated with the fact that a long-term influence of the
Galactic bar led to a characteristic bimodal splitting of the $UV$
velocity plane.

 \thanks{
 This study was supported by the ``Transient and explosive processes in Astrophysics''
 Program of the Presidium of Russian Academy of Sciences (P--7).
 }

 \References
  \refb Allen C. and  Santill\'an A., 1991, Rev. Mex. Astron. Astrof., 22, 255
  \refb Antoja T., Figueras F., Fern\'andez D., et al., 2008, A\&A, 490, 135
  \refb Asiain R., Figueras F.,  Torra J., et al., 1999,  A\&A, 341, 427
  \refb Bensby T., Oey M. S., Feltzing S., et al., 2007, ApJ, 655, L89
  \refb Bobylev V. V., Gontcharov G. A., Bajkova A. T., 2006, Astron. Rep., 50, 733
  \refb Bobylev V. V., Bajkova A. T., 2007, Astron. Rep., 51, 372
  \refb Bobylev V. V., Bajkova A. T.,  Myll\"ari A. A., 2010, Astron. Lett., 36, 27
  \refb Bobylev V. V., Bajkova A. T., 2016, Astron. Lett., 42, 228
  \refb Chakrabarty D., 2007, A\&A, 467, 145
  \refb Chereul E., Cr\'ez\'e M.,  Bienaym\'e O., 1998, A\&A, 340, 384
  \refb Dehnen W., 1998, AJ, 115, 2384
  \refb Dehnen W., 1999, ApJ, 524, L35
  \refb Dehnen W., 2000, AJ, 119, 800
  \refb Famaey B., Jorissen A., Luri X., et al., 2005, A\&A, 430, 165
  \refb Fern\'andez D., Figueras F., Torra J., 2008, A\&A, 480, 735
  \refb Francis C., Anderson E., 2009, New Astron., 14, 615
  \refb Fux R., 2001, A\&A, 373, 511
  \refb Gardner E., Flinn C., 2010, MNRAS, 405, 545
  \refb Gontcharov G. A., 2006, Astron. Lett. 32, 795
  \refb The Hipparcos and Tycho Catalogues, 1997, ESA SP--1200
  \refb Jung C., Zotos E. E., 2016, MNRAS, 457, 2583
  \refb van Leeuwen F., 2007, A\&A, 474, 653
  \refb Palou$\breve{s}$ J., Jungwiert B., Kopeck$\acute{y}$ J., 1993, A\&A, 274, 189
  \refb Sch\"onrich R., Binney J., Dehnen W., 2010, MNRAS, 403, 1829
  \refb Skuljan J., Hearnshaw J. B., Cottrell P. L., 1999, MNRAS, 308, 731

 \end{document}